\documentclass[showpacs,twocolumn,amsmath,amssymb,superscriptaddress]{revtex4-1}
\usepackage{amsmath,amssymb}
\usepackage{bbm}
\usepackage{graphicx,graphics}
\usepackage{amsmath,amsfonts,amsthm,bm}

\begin{document}

\title{Revealing excess protons in the infrared spectrum of liquid water}
\author{V. G. Artemov}\email{Corresponding author: v.artemov@skoltech.ru}
\affiliation{Center for Energy Science and Technology, Skolkovo Institute of Science and Technology, 3 Nobel Street, Moscow 121205, Russia}
\affiliation{1. Physikalisches Institut, Universit\"at Stuttgart, 70569 Stuttgart, Germany}
\author{E. Uykur}
\affiliation{1. Physikalisches Institut, Universit\"at Stuttgart, 70569 Stuttgart, Germany}
\author{S. Roh}
\affiliation{1. Physikalisches Institut, Universit\"at Stuttgart, 70569 Stuttgart, Germany}
\author{A. V. Pronin}
\affiliation{1. Physikalisches Institut, Universit\"at Stuttgart, 70569 Stuttgart, Germany}
\author{H. Ouerdane}
\affiliation{Center for Energy Science and Technology, Skolkovo Institute of Science and Technology, 3 Nobel Street, Moscow 121205, Russia}
\author{M. Dressel}
\affiliation{1. Physikalisches Institut, Universit\"at Stuttgart, 70569 Stuttgart, Germany}

\begin{abstract}
The most common species in liquid water, next to neutral H$_2$O molecules, are the H$_3$O$^+$ and OH$^-$ ions. In a dynamic picture, their exact concentrations depend on the time scale at which these are probed. Here, using a spectral-weight analysis, we experimentally resolve the fingerprints of the elusive fluctuations-born short-living H$_3$O$^+$, DH$_2$O$^+$, HD$_2$O$^+$, and D$_3$O$^+$ ions in the IR spectra of light (H$_2$O), heavy (D$_2$O), and semi-heavy (HDO) water. We find that short-living ions, with concentrations reaching $\sim 2\%$ of the content of water molecules, coexist with long-living pH-active ions on the picosecond timescale, thus making liquid water an effective ionic liquid in femtochemistry.
\end{abstract}
\date{\today}

\keywords{}

\maketitle

\section*{Introduction}
H$_3$O$^+$ and OH$^-$ ions, characterized by picosecond-scale lifetimes in liquid water \cite{ref1}, are important species in biology, chemical physics, and electrochemistry, as they drive the charge transfer process and acid-base reactions in aqueous solutions, and serve as intermediates for protonic transport \cite{ref2,ref3}. Their study dates back as early as 1806 when Grotthuss published his far-sighted theory of water conductivity \cite{ref3a}. Since then hydronium and hydroxide ions have been intensively investigated, yet their dynamical properties (i.e., the faster, femtosecond-scale, fluctuations) remain elusive though, thus precluding the detailed understanding of the exact ions' role in the microscopic properties of liquid water. 

The presence of ionic species in pure water has been confirmed by electrochemical measurements, and their infrared (IR) and Raman spectral signatures have been found experimentally in aqueous solutions and hydrated hydronium clusters \cite{ref3b,ref3c}; however the assignment of the water ions to actual (observable) spectral features of pure water are, in fact, still crucially missing \cite{ref4,ref5,ref6,ref7}. For example, nuclear magnetic resonance, potentiometry, and conductometry are low-speed methods and hence provide only limited information on water's microscopic properties. The faster techniques, e.g., pump-probe \cite{ref8}, terahertz \cite{ref9}, and IR \cite{ref10} spectroscopy are better suited to probe ultra-fast (sub-picosecond) dynamics, but miss the characteristic features of the water ions because of their low concentrations compared to that of the neutral H$_2$O molecules, causing the ions' spectral signatures to be hidden \cite{ref11}. The dynamics of water ions has been intensively studied by theory, and particularly important results have been obtained by numerical simulations \cite{ref11b,ref3aa}. 

Femtosecond fluctuations of molecules, causing the proton transfer reactions \cite{ref11c}, lead to the spontaneous formation of short-living (SL, denoted with *) ions, H$_3$O$^{+*}$ and OH$^{+*}$ \cite{ref12}, which differ from the ``conventional'' long-living (LL) ions, as only the latter contribute to pH and are detectable with slow measurement techniques. The formation-recombination pathway of SL ions is intermolecular and loop-shaped \cite{ref13}, and it yields ``instantaneous'' SL concentrations much higher than LL's ones \cite{ref16,ref17,ref18}, but random thermal disturbances allow transfer of SL ions to the LL pH-active H$_3$O$^+$ and OH$^-$ ion populations \cite{ref14,ref15}. Importantly, because of their short lifetime ($\sim$ 3 ps) SL ions contribute only little to dc conductivity \cite{ref19,ref20}, yet their concentration can be sufficiently large ($\sim$1 M) to drive the Debye relaxation of water \cite{ref18,ref19}. 

Since on time scales shorter than picoseconds, water's chemical composition can be described in terms of ionic and molecular species, both LL and SL ions should contribute to the IR (and higher energy) spectrum. However, the structure of liquid water IR spectrum is such that it cannot be used for the direct identification of the ions' contributions because of strongly overlapping bands with multiple contributions, as demonstrated in Ref.~\cite{ref21}. Moreover, ionic vibrations, which are absent in water vapor \cite{ref22} and cannot be easily identified in the condensed phase, are entangled with intramolecular oscillations in liquid water \cite{ref23}. Interpretations of liquid water's IR spectrum up to now available hence appear to necessarily rely on sophisticated analyses and models; here, following the \emph{lex parsimoniae} principle, we attempt to derive a simple interpretation to provide a straightforward, more transparent physical picture. 

In this work, we perform Fourier transform IR spectroscopy experiments of liquid light, semi-heavy, and heavy water; we analyze the data obtained with a model of the dynamical structure of liquid water, to elucidate the elusive ionic contributions to its IR spectrum. In comparison with previous studies \cite{ref3b,ref3c}, we study water itself, rather than aqueous solutions or other multicomponent systems, to detect the ionic species in the natural bulk water environment. The replacement of protons (one or several) in the ionic species by deuterons, shifts spectroscopic lines and makes ionic species ``visible'' in the integral spectrum. The spectral-weight analysis shows that up to picosecond timescale water is composed of a significant amount of short-living ionic species, influencing the nature of intermolecular interactions on this ultra-short time scale.

\section*{Results and analysis}

\subsection*{Spectra}

\begin{figure}[h!]
\includegraphics[width=0.5\textwidth]{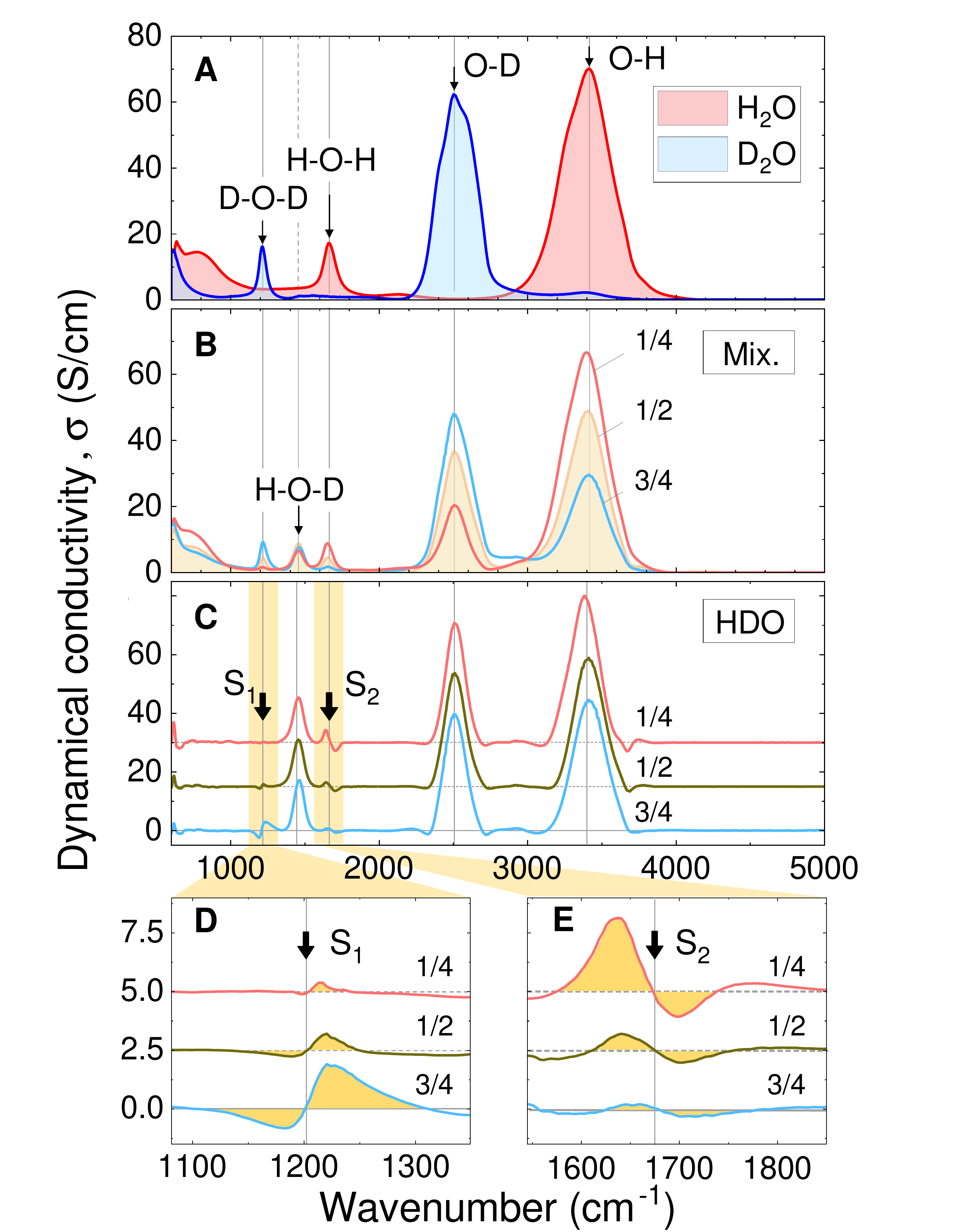}
\caption{Experimental spectra of dynamical conductivity $\sigma(\nu)$ of (A) pure light and pure heavy water, (B) mixtures of light and heavy water with different molar isotopic fractions $f$ as indicated, (C) pure HDO calculated using Eq.~\eqref{sigmix}, and (D,E) magnified parts of the HDO spectra near the D$-$O$-$D and H$-$O$-$H bending modes showing ``S-shaped'' mismatches (see text). The spectra for $f=1/2$ and $f=1/4$ in panels (C-E) are shifted upwards for clarity. Thin arrows show two stretching modes above 2000 cm$^{-1}$, and three bending modes below 2000 cm$^{-1}$. Vertical lines are guides for the eye. Thick arrows indicate the areas, where the mismatches of the spectra, $S_1$ and $S_2$, occur.
}\label{fig1}
\end{figure}

Figure~\ref{fig1}A,B shows the experimental transmission infrared (TIR) spectra of dynamical conductivity $\sigma(\nu)$ for the water mixtures with different molar fractions, $f = D/(H+D)$, of heavy (D$_2$O) water ($0 \leq f \leq 1$).  Here, $D$ and $H$ represent the concentrations of deuterons and protons, respectively. Details of the measurement procedure are given in the Methods section. Analyzing the absorption spectra in terms of dynamical conductivity has an important advantage: the integral under a mode in the $\sigma(\nu)$ curve is directly proportional to the number of charges involved in the absorption mechanism responsible for this mode \cite{ref32, ref32a}, e.g., in corresponding vibration or rotation. This allows one to account for the concentration of species that contribute to the certain mode. One can clearly see the two groups of vibrational bands: at 1000-2000 and 2000-4000 cm$^{-1}$. The central frequencies and integral intensities of these bands are presented in Table~\ref{tab1}. The right-side portion of the spectrum (frequencies above 2000 cm$^{-1}$) contains the O$-$H and O$-$D stretching modes, while the left-side portion (below 2000 cm$^{-1}$) hosts the H$-$O$-$H, D$-$O$-$D, and H$-$O$-$D bending modes \cite{ref27}. Obviously, the H$-$O$-$D modes appear only in spectra of H$_2$O/D$_2$O mixtures (see Fig.~\ref{fig1}B). All bands are intense, wide, and possess complex internal structures (consisting of several Lorentzians). The relative shift of the H$_2$O and D$_2$O lines well matches the shift of the normal modes of the corresponding harmonic oscillators (see Appendix). 

\begin{table}
\caption{\label{tab1}Frequencies $\nu$ in cm$^{-1}$ (upper rows) and integral intensities $I$ in S$\cdot$cm$^{-2}$ (in brackets on lower rows) of the bending and stretching vibrations peaks shown in Fig.1.}
\begin{tabular}{c|ccccc}
$f$ & DOD & HOD & HOH & OD & OH\\
\hline
0 & n/a & n/a & 1658 & n/a & 3414\\
~ & ~ & ~ & (3360) & ~ & (29745)\\
~ & ~ & ~ & ~ & ~ & ~\\
1/4 & 1217 & 1461 & 1651 & 2510 & 3413\\
~ & (216)	& (789)	& (916) & (4800) & (24552)\\
~ & ~ & ~ & ~ & ~ & ~\\
1/2 & 1217 & 1457 & 1653 & 2509 & 3410\\
~ & (385)	& (961)	& (509) & (8913) & (16822)\\
~ & ~ & ~ & ~ & ~ & ~\\
3/4 & 1214 & 1458 & 1651 & 2507 & 3396\\
~ & (735) & (793) & (215) & (13303) & (10023)\\
~ & ~ & ~ & ~ & ~ & ~\\
1 & 1213 & n/a & n/a & 2506 & n/a\\
~ & (1554) & ~ & ~ & (19351) & ~\\
\end{tabular}
\end{table}

While the spectra of light and heavy water can easily be obtained in an experiment, the ``actual'' spectrum of HDO cannot be measured directly. In order to pinpoint the HDO modes, spectra of the mixtures can be analyzed as weighted sums of the spectra of light, heavy, and semi-heavy water \cite{ref28,ref29}; hence, we may write the dynamical conductivity spectra of mixtures $\sigma_{\rm mix}(\nu,f)$ with different $f$ as:
\begin{equation}\label{sigmix}
  \sigma_{\rm mix}(\nu,f) = a\sigma_{\rm D_2O}(\nu) + b\sigma_{\rm H_2O}(\nu) + c\sigma_{\rm HDO}(\nu)
\end{equation}

\noindent where $\sigma_{\rm D_2O}(\nu)$, $\sigma_{\rm H_2O}(\nu)$, $\sigma_{\rm HDO}(\nu)$ are $f$-independent spectra, and the coefficients $a=f^2$, $b=(1-f)^2$, and $c=2f(1-f)$, are the probabilities of formation of D$_2$O, H$_2$O, and HDO, respectively. As the zero point energy difference between H and D leaves the \emph{bending} coordinate unchanged from the normal coordinate \cite{ref29b}, the HDO spectrum calculated from Eq.~\eqref{sigmix} should be $f$-independent.

Figure~\ref{fig1}C shows spectra of HDO calculated using Eq.~\eqref{sigmix} from the spectra displayed in Fig.~\ref{fig1}A,B. Hereinafter, we concentrate entirely on the analysis of the bending modes to avoid any influence of Fermi resonances, which are known to affect the higher frequency stretching vibrations, but are not expected to provide any significant distortions to the bending modes as these modes are those with the lowest eigenfrequencies among all vibrations. We note that the processes affecting the shapes of the bending modes and discussed in detail below, can also affect the shapes of the stretching modes. However, we do not discuss the stretching modes here, because the point of our study is not to precisely describe the entire spectrum of water, but to use this spectrum as a tool in determining the content of different ions on fast time scales. For this purpose, the bending modes are much better suited.

The HDO bending spectra generally coincide with each other; however, there are very characteristic mismatches (S-shaped features) near the D$-$O$-$D and H$-$O$-$H bands (Fig.~\ref{fig1}C). The presence of these S-features suggests either that the measured spectra are not accurate enough, species other than H$_2$O, D$_2$O, and HDO, are present in the mixture, or substitution of H$_2$O by D$_2$O affects anharmonic resonances. We are inclined to consider the second scenario for the following reasons. First, we found that in our measurements, the intensities of the mismatches shown in Fig.~\ref{fig1}C surely exceed the uncertainty of the transmission mode (TIR) measurements and the error of spectra processing (see Appendix). Next, very similar S-shape mismatches in the HDO spectra obtained in the same frequency region have been reported earlier \cite{ref28,ref29}. Most importantly, the frequency, and amplitude of the S-features are reproducible and consistently changing with the molar fraction $f$. Furthermore, the S-feature near the D$-$O$-$D bend (Fig.~\ref{fig1}D) is mirror-symmetric in shape with respect to the S-feature near the H$-$O$-$H bending mode (Fig.~\ref{fig1}E). This observation is in contrast to one of the previous studies \cite{ref28}, where both S-features are shown identical; but it is in agreement with a more precise recent study \cite{ref29}, where the similar pattern of the S-features was demonstrated in the analysis of absorption spectrum. Finally, the ratio between the integral intensities of the positive and negative components for the S-feature (see Figs.~\ref{fig1}D,E) is 3:1, independently on the molar fraction $f$. From these points we conclude that the S-features are due to internal species of liquid water that are not taken into account by Eq.~\eqref{sigmix}, as this relation includes only molecular components. Below we argue that short-living intrinsic ions of water are good candidates for the missing water species and that the density of these ions is relatively high.

\subsection*{Chemical composition of liquid water}
Fig. 2A depicts nine principal species - neutral molecules and ions - for mixtures of light and heavy water. The species H$_2$O, H$_3$O$^{+*}$, OH$^{-*}$ (group 1) and their deuterated isotopologues D$_2$O, D$_3$O$^{+*}$, OD$^{-*}$ (group 2) constitute pure light and heavy water, respectively. The species HDO, HD$_2$O$^{+*}$, DH$_2$O$^{+*}$ (group 3) are only present in water mixtures, where they coexist with the six species of the groups 1 and 2. Thus, in addition to the three molecular species (line i of Fig. 2A), a mixture contains also four positively charged ions (line ii) and two negatively charged ions (line iii). The relative proportion of these nine species is determined by the autoionization-recombination events, but it also depends on the molar fraction $f$. Here and below, we do not distinguish the SL ions and LL ions: albeit their lifetimes differ, they are indistinguishable on the short timescales relevant for IR vibrations. 

\begin{figure}[h!]
\includegraphics[width=0.5\textwidth]{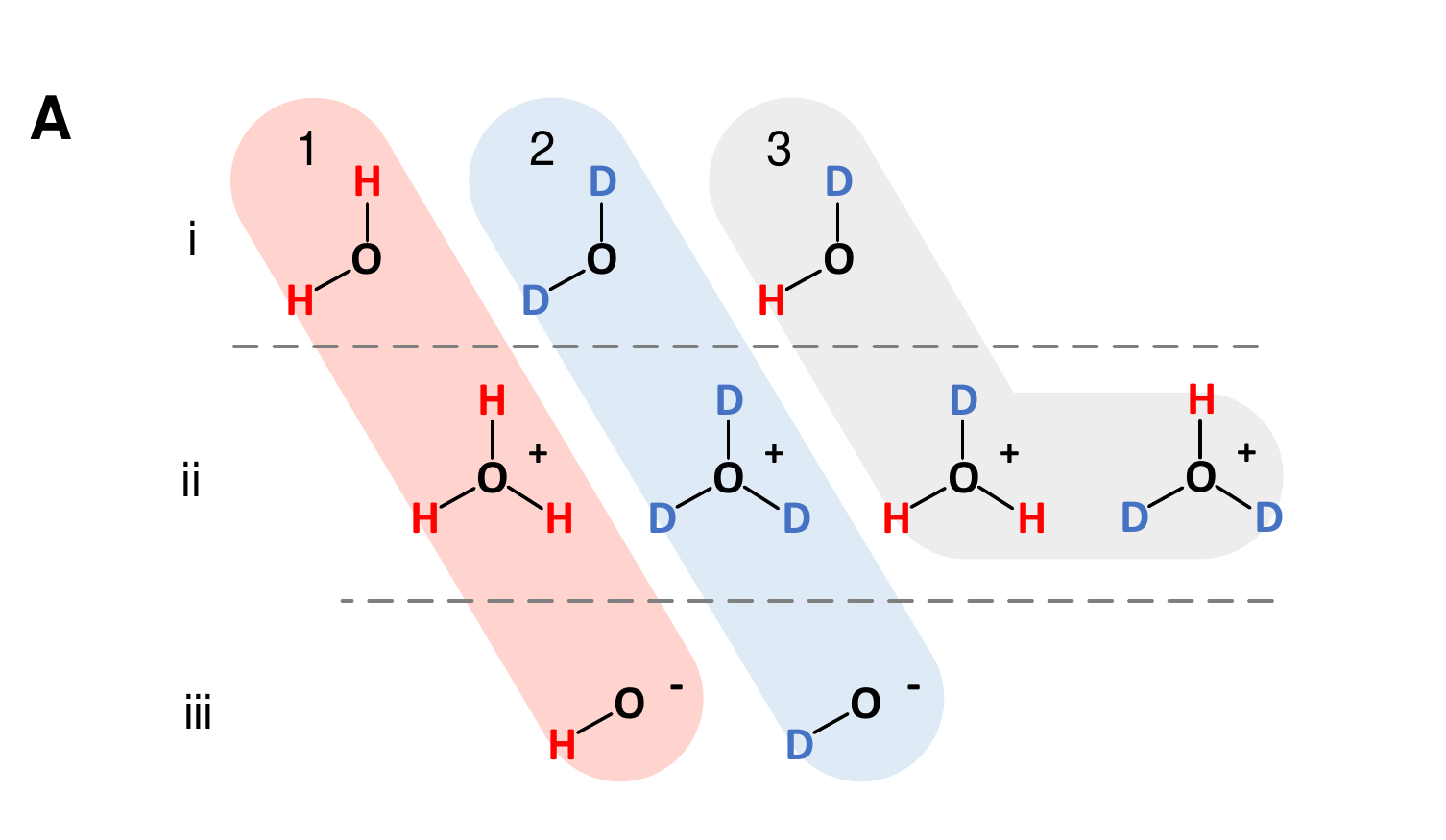}
\includegraphics[width=0.5\textwidth]{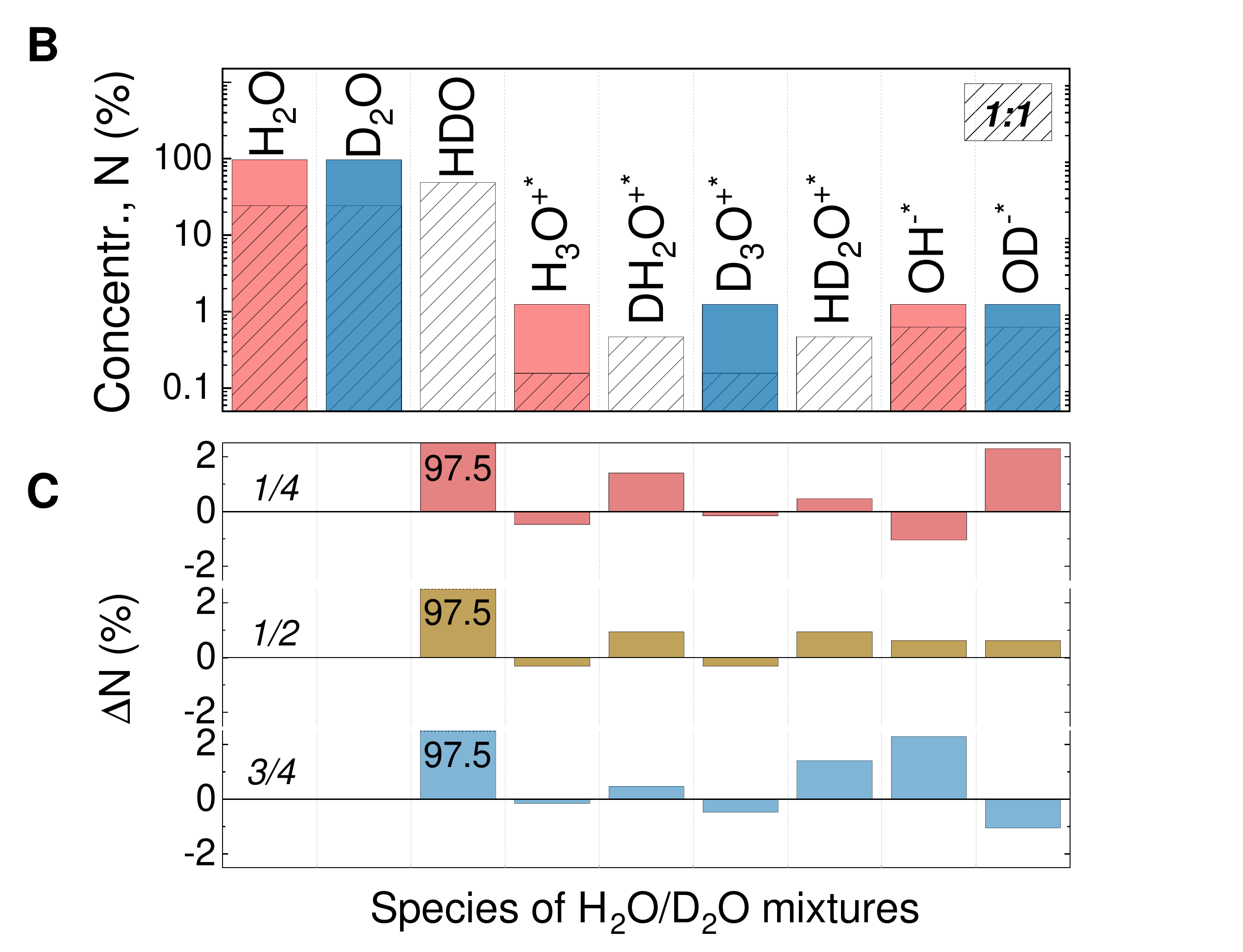}
\caption{(A) Nine principal water species, present in light and heavy water (groups 1 and 2, respectively), and in their mixtures (all three groups). The rows correspond to neutral molecules (row i) and positively (row ii) and negatively (row iii) charged ions. (B) Percentage of the water species in pure H$_2$O (red), pure D$_2$O (blue) and their 1:1 mixture (shaded) according to Table~\ref{tab2}; note the logarithmic scale on the $y$-axis. (C) Differences in ion concentrations obtained after subtraction of the normalized bars for pure light and heavy water (red and blue) from the ones for the mixtures (shaded), according to Eq.(\ref{sigmix}). Numbers near zero-lines are the molar fractions $f$. The data shown in panel (B) correspond to the middle raw in panel (C).}\label{fig3}
\end{figure}

Inasmuch as it is difficult to detect concentration the SL ions in water directly, we rely on studies based on the terahertz absorption measurements \cite{ref17}, as well as on molecular dynamics simulations that take into account the auto-dissociation events in the frame of the semi-classical approach \cite{ref20}. Both studies conclude that the approximate concentration of the SL ions is about 1 M (or about 2\%). In a more recent investigation \cite{ref18} - based on the analysis of broadband spectrum of water - the concentration of ionic species accounts to 2.5\%. Hereinafter we choose this value for our preliminary analysis, and it will be later confirmed by the fit of the experimental data.

\begin{table}
\caption{\label{tab2}Percentage of species (neutral molecules and SL ions) in liquid H$_2$O, D$_2$O and their mixtures.}
\begin{tabular}{c|ccccc}
$f$ & 0 & 1/4 & 1/2 & 3/4 & 1\\
\hline
HDO & 0	& 36.56	& 48.75	& 36.56	& 0\\
~ & ~ & ~ & ~ & ~ & ~\\
H$_2$O & 97.5	& 54.84	& 24.38	&6.094 & 0\\
~ & ~ & ~ & ~ & ~ & ~\\
D$_2$O & 0 & 6.094 & 24.38 & 54.84 & 97.5\\
~ & ~ & ~ & ~ & ~ & ~\\
H$_3$O$^{+*}$ & 1.25	& 0.527	& 0.156	& 0.020	& 0\\
~ & ~ & ~ & ~ & ~ & ~\\
D$_3$O$^{+*}$ & 0 & 0.020 & 0.156 & 0.527 &1.25\\
~ & ~ & ~ & ~ & ~ & ~\\
DH$_2$O$^{+*}$ & 0	& 0.527	& 0.469	& 0.176	& 0\\
~ & ~ & ~ & ~ & ~ & ~\\
HD$_2$O$^{+*}$ & 0	& 0.176	& 0.469	& 0.527	& 0\\
~ & ~ & ~ & ~ & ~ & ~\\
OH$^{-*}$ & 1.25	& 0.313	& 0.625	& 0.938	& 0\\
~ & ~ & ~ & ~ & ~ & ~\\
OD$^{-*}$ & 0 & 0.938 & 0.625 & 0.313 & 1.25\\
\end{tabular}
\end{table}

Table \ref{tab2} lists the calculated percentage of nine principal species (``particles'') of water mixtures for different molar fractions $f$. The concentrations of the particles are obtained by the combinatorial probability of permutation of H and D atoms around the oxygen atom O in the assumption that: (i) the probabilities for formation of the covalent H and D bonds are the same; (ii) the relative percentages of ions and molecular species are identical for all mixtures; (iii) the SL ions concentration is $n_i = 2.5$\%; and (iv) the concentration of positively charged ions is equal to the concentration of negatively charged ions, $n^+=n^-=n_i/2$ (charge-neutrality). The concentrations of positively charged ions are calculated from the total concentration $n^+$ of positive ions using the following coefficients $\alpha$=$(1-f)^3$, $\beta$=$f^3$, $\gamma$=$3f(1-f)^2$, and $\delta$=$3f^2(1-f)$ for H$_3$O$^{+*}$, D$_3$O$^{+*}$, DH$_2$O$^{+*}$, and HD$_2$O$^{+*}$, respectively. These coefficients are the probabilities to form these ions from H and D atoms available in the mixtures. We obtained the concentrations of negatively charged ions $n^-$, using $f$ and $(1-f)$ as probability coefficients for OH$^{-*}$ and OD$^{-*}$, respectively.

Figure \ref{fig3}B presents the concentration bars for all 9 possible species in light water ($f$ = 0), heavy water ($f$ = 1) and their 1:1 mixture ($f$ = 1/2). The column bars correspond to the values given in Table \ref{tab2}. As long as the total concentrations of positive and negative ions (charges) $n^+$ and $n^-$ are fixed, they are distributed among a larger number of different species (types of ions) in a mixture than in pure H$_2$O or D$_2$O. For example, the population density of OH$^{-*}$ of pure light water in the mixture is distributed between OH$^{-*}$ and OD$^{-*}$. The same for positively charged species: the asymmetric HD$_2$O$^{+*}$ and DH$_2$O$^{+*}$ ions, present in the mixtures only, take a part of the total charge density from the symmetric H$_3$O$^{+*}$ and D$_3$O$^{+*}$ ions (see the differences between the bars for pure H$_2$O or D$_2$O and for the mixtures in Fig.\ref{fig3}B). This effect was unaccounted for in the above IR-spectrum analysis. 

When we apply the normalization coefficients $a$, $b$, and $c$ (the same as in Eq.~\eqref{sigmix}) to the column bars of the pure and the mixed H$_2$O/D$_2$O, only the concentrations of \emph{molecular} species are leveled off. For the ionic species, this is not the case: their bars (in Fig. 2B) for the pure probes and for the mixtures do not coincide. This is because the relative ratio of the ionic species and the molecular species in any mixture differs from those in pure light or heavy water. In other words, this simple normalization-subtraction procedure properly works for the molecular species only, while for the ionic species the presence of additional spectral contributions is now apparent.

Figure \ref{fig3}C exhibits the differences obtained after subtraction of the normalized bars for pure light and heavy water from the bars corresponding to the mixtures (for $f = 1/4,~ 1/2,~ \mbox{and}~ 3/4$). As one can see, the H$_2$O and D$_2$O bars are subtracted completely, and all evaluated HDO-concentration columns coincide with each other ($\Delta$N=97.5\%), but the ionic bars differ. The number of H$_3$O$^{+*}$ and D$_3$O$^{+*}$ ions and the number of HD$_2$O$^{+*}$ and DH$_2$O$^{+*}$ ions are, respectively, over- and under-estimated. Asymmetric ions are not subtracted because they do not exist in pure light and heavy water, while concentrations of symmetric ions are over-evaluated. The same applies for negatively charged OH$^{-*}$ and OD$^{-*}$ ions. Their relative concentrations in the mixture differ from those for neutral molecules, which leads to the corresponding residual bars after subtraction.

As the probability of the formation of asymmetric ions is three-times-higher than for symmetric ones, the relative ratio of the concentrations of up-going and down-going residual bars of positively charged ions in Fig.~\ref{fig3}C is 3:1 everywhere (compare the bars for H$_3$O$^{+*}$ and DH$_2$O$^{+*}$, as well as, D$_3$O$^{+*}$ and HD$_2$O$^{+*}$ for each probe). The ratio is the same as observed for the positive and negative parts of the experimentally observed S-features in Fig. \ref{fig3}. This, as well as the fact that the intensities of S-features are proportional to the concentration of ionic species, make it possible to identify ions in the spectrum and calculate their concentrations.

\subsection*{Assignment of the S-features}
Because the OH$^{-*}$- and OD$^{-*}$-ion vibrations are very close to the O$-$H and O$-$D stretching modes, they are hidden within the bands at 2450 and 3400 cm$^{-1}$ (see Fig.~\ref{fig1}C). However, the positively charged ions, D$_3$O$^{+*}$, HD$_2$O$^{+*}$, DH$_2$O$^{+*}$, and H$_3$O$^{+*}$ possess vibrational modes, whose frequencies are different from each other, and also from the HDO, H$_2$O, and D$_2$O frequencies. Thus, the concentrations of the positively charged ions in the mixtures can be identified by comparing their IR spectra.

\begin{figure}[h!]
\includegraphics[width=0.5\textwidth]{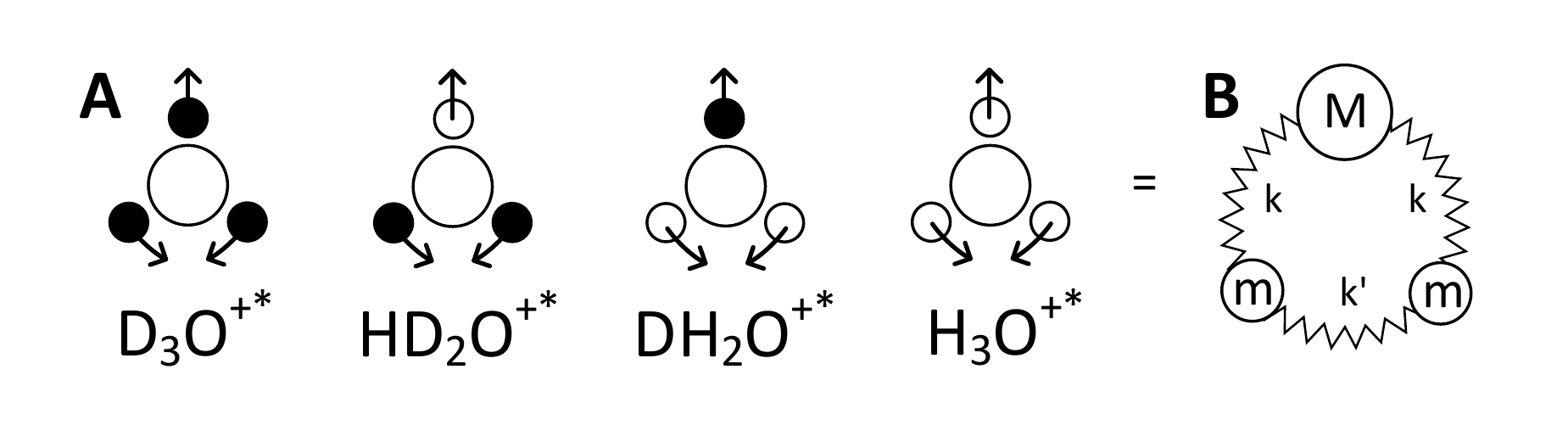}
\caption{(A) Bending oscillations of positively charged ions in mixtures of light and heavy water. Black circles are D atoms, small open circles are H atom, large open circles are O atoms. (B) A harmonic-oscillator model representing these vibrations with the normal modes defined by Eq.~\eqref{eigenf}.}\label{fig4}
\end{figure}

Figure~\ref{fig4} sketches the bending vibrations of positively charged ions, whose frequencies are expected to be close to (but still different from) the bending frequencies of H$_2$O and D$_2$O at 1650 and 1240 cm$^{-1}$, respectively (see Fig.~\ref{fig1}). If we now utilize a simple harmonic-oscillator model with three masses and three springs depicted in Fig.~\ref{fig4}, two modes with the following nonzero eigenfrequencies are expected:

\begin{equation}\label{eigenf}
\omega_1 = \sqrt{(2k' +k)/m}~~\mbox{and}~~ \omega_2 = \sqrt{2k/\mu}
\end{equation}

\noindent where $k$ and $k'$ are the spring constants, $\mu=2mM/(2m+M)$ is the effective mass (which is given in Table~\ref{tab3} for different species of water), $m$ is the mass of deuteron or proton, and $M$ is the mass of one oxygen atom (plus the mass of one deuteron or proton in the case of DH$_2$O$^{+*}$ and HD$_2$O$^{+*}$ ions). The mode with the eigenfrequency $\omega_2$ corresponds to bending vibrations shown in Fig.~\ref{fig4}. The second mode is a stretching vibration and is not relevant for this study. The spring constant $k'$ (which correspond to the vibrations between the light atoms) is close to zero, while $k$ (the vibrations between a light atom and heavy oxygen) is equal to $142 \pm 4$ N/m for both H$_2$O and D$_2$O, as can be defined from the experimental frequencies of H$_2$O and D$_2$O molecules and Eq.~\eqref{eigenf} \cite{ref30}. 

\begin{table}
\caption{\label{tab3}Parameters of oscillations shown in Fig.~\ref{fig4} and bending normal-mode frequencies, $\nu_2=\omega_2/2\pi$, calculated with Eq.~\eqref{eigenf}, comparison with the experimental results, $\nu_2^{\rm exp}$. }
\begin{tabular}{c|cccccc}
~ & $M$ & $m$ & $\mu$ & $k$ & $\nu_2$ & $\nu_2^{\rm exp}$\\
~ & (Da) & (Da) & (Da) & (N$\cdot$m$^{-1}$) & (cm$^{-1}$) & (cm$^{-1}$)\\
\hline
H$_2$O & 16	& 1	& 1.778	& 145	& 1663 & 1638\footnotemark[1]\\
~ & ~ & ~ & ~ & ~ & ~ &\\
D$_2$O & 16	& 2	& 3.200	& 139	& 1213 & 1205\footnotemark[2]\\
~ & ~ & ~ & ~ & ~ & ~ &\\
H$_3$O$^{+*}$ & 17	& 1	& 1.789	& 151	& 1695 & 1718\footnotemark[3]\\
~ & ~ & ~ & ~ & ~ & ~ &\\
DH$_2$O$^{+*}$ & 18 & 1 & 1.800 & 142 & 1637	& no data\\
~ & ~ & ~ & ~ & ~ & ~ &\\
HD$_2$O$^{+*}$ & 17 & 2 & 3.238 & 142 & 1222 & no data\\
~ & ~ & ~ & ~ & ~ & ~ &\\
D$_3$O$^{+*}$ & 18	& 2	& 3.273	& 138	& 1195 & 1194\footnotemark[1]\\
~ & ~ & ~ & ~ & ~ & ~ &\\
\end{tabular}
\footnotetext[1]{From Ref.~\cite{Fagiani2016}.}
\footnotetext[2]{From Ref.~\cite{ref26}.}
\footnotetext[3]{From Ref.~\cite{ref27}.}
\end{table}

The constant $k$ is not expected to change much across the isotopologues and from Eq.~\eqref{eigenf} one should expect that: (i) the heavier ions have lower oscillation frequencies and (ii) D$_3$O$^{+*}$ and HD$_2$O$^{+*}$ have their vibrational frequencies close to the D$_2$O bending mode, while the frequencies of DH$_2$O$^{+*}$ and H$_3$O$^{+*}$ are close to the bending mode of H$_2$O. Thus, the spectral fingerprints of the ions illustrated in Fig.~\ref{fig4} are expected to appear in the following order with decreasing frequency: H$_3$O$^{+*}$, DH$_2$O$^{+*}$, HD$_2$O$^{+*}$, and D$_3$O$^{+*}$, first two being around 1660 cm$^{-1}$, while the last two showing up near 1210 cm$^{-1}$.

\begin{figure}[h!]
\includegraphics[width=0.5\textwidth]{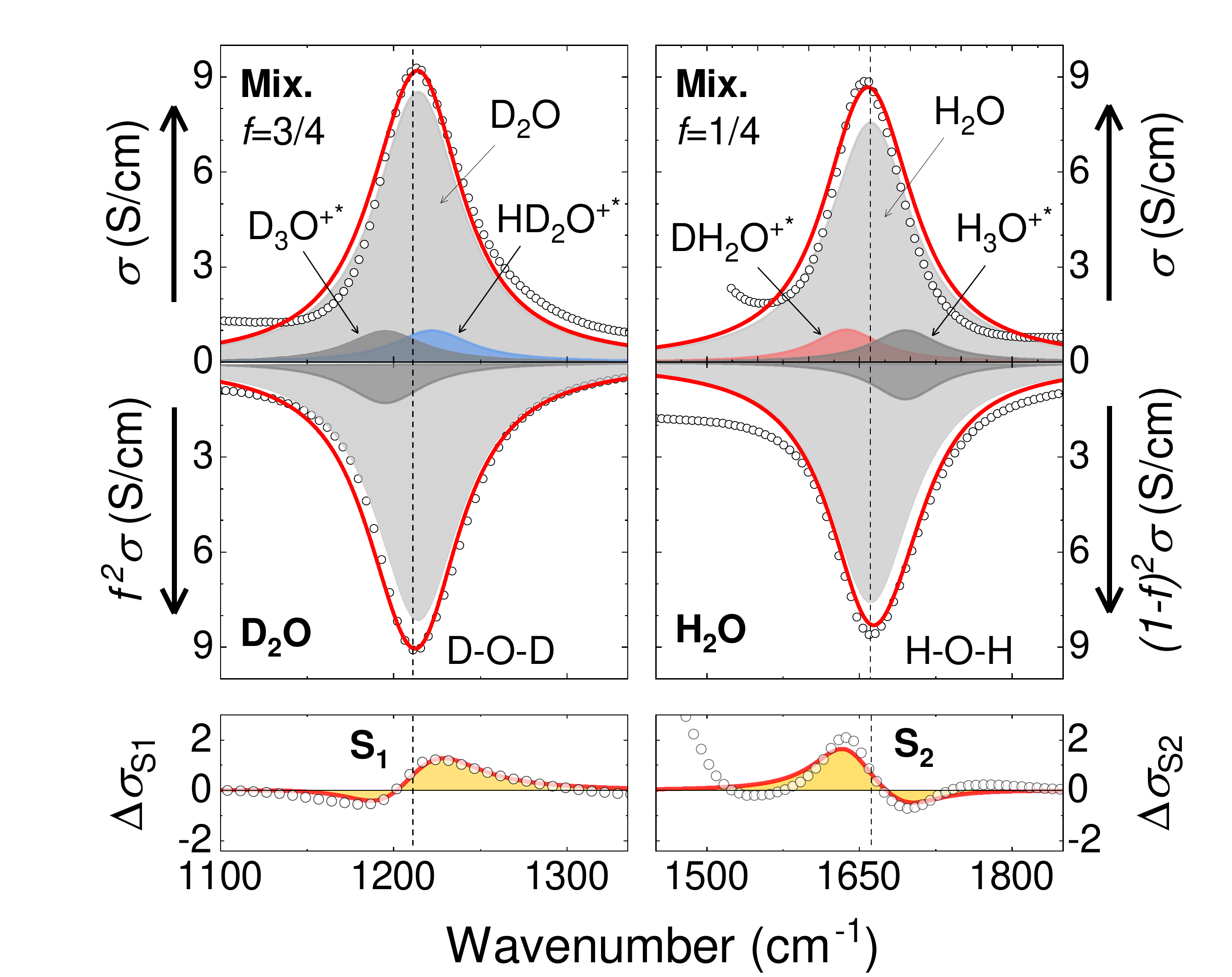}
\caption{Examples of the S-feature fits near the bending modes. The total fits are shown as red lines, while different Lorentzian terms are represented as shaded areas. The experimental data are open symbols. Left (right) panels display the spectra near the resonances at around 1210 (1660) cm$^{-1}$. The fits are shown for the largest observed S-features (cf. Fig.~\ref{fig1}D,E). The top row of panels shows the conductivity spectra for the $f$=3/4 (left) and $f$=1/4 (right) mixtures, respectively. In the middle row, the conductivities for pure D$_2$O (left) and H$_2$O (right) are shown upside down to demonstrate the subtraction procedure utilized for the S-feature fits, as described by Eqs.~\eqref{ds1} and ~\eqref{ds2}. The resultant S-feature fits are shown in the bottom panels.}\label{fig5}
\end{figure}

We now provide the parameterization of the observed S-features. Figure~\ref{fig5} explains graphically how the fits of these features were obtained. The Lorentz-oscillator parameters for the ions were acquired by performing a simultaneous fitting of the measured spectra and the corresponding S-features from Fig.~\ref{fig1} using the following equations:

\begin{eqnarray}
\nonumber
\label{ds1}
\Delta\sigma_{\rm {S_1}} & = &  [f^2(\sigma_{\rm D_2O}+\beta \sigma_{\rm D_3O^{+*}})+2f(1-f)\delta \sigma_{\rm HD_2O^{+*}}]\\ 
&&- f^2\left[\sigma_{\rm D_2O}+\sigma_{\rm D_3O^{+*}}\right]\\
\nonumber
\label{ds2}
\Delta\sigma_{\rm {S_2}} & = & [(1-f)^2(\sigma_{\rm H_2O}+\alpha \sigma_{\rm H_3O^{+*}})+2f(1-f)\gamma \sigma_{\rm DH_2O^{+*}}]\\ 
&&- (1-f)^2[\sigma_{\rm H_2O}+\sigma_{\rm H_3O^{+*}}]
\end{eqnarray}

\noindent where $\alpha$, $\beta$, $\gamma$, $\delta$ are the same as defined above, and  all $\sigma_i$'s are functions of $\nu$: $\sigma_i(\nu) = I_iw_i^2/\left[w_i^2+(\nu_{{\rm o}i}-\nu)^2\right]$ is the Lorentz-oscillator contribution to the dynamical conductivity, with $I_i$ being its intensity, $w_i$ its half-width, $\nu_{{\rm o}i}$ the oscillator central frequency, and $f$ is the D$_2$O molar fraction in the water mixture, as already defined. These formulas follow from Eq.~\eqref{sigmix}, if the water mixtures contain ions with the concentrations defined above (see the section “Chemical composition of liquid water”) and if we apply the spectra subtraction procedure, discussed above for the experimentally obtained S-features, to our fits. The fit parameters are given in Table~\ref{tab4}. One can see a very good agreement between the model and the experimental data. Thus, the mismatches in the subtracted HDO spectra are assigned to the SL ionic species that are parts of the IR spectra of light and heavy water, as well as their mixtures.

S-features in the subtracted spectra of HDO  were already reported in \cite{ref29}, where spectral data were analyzed in terms of absorption $\alpha=4\pi\sigma(\nu)/n(\nu)c$, where $c$ is the speed of light, $n(\nu)$ is the frequency-dependent refractive index; unfortunately it cannot reflect the true concentration of ionic species. Nevertheless, the residues in the HDO spectra are quite similar to ours, indirectly confirming our results. The authors interpret the residues in terms of hydrogen bonding, considering the perturbation of the intramolecular vibrations caused by surroundings, as they assumed that molecular species (H$_2$O, D$_2$O, and HDO) oscillation frequencies change by the neighboring atoms that provide hydrogen or deuterium bonds \cite{ref29}. In spite of their very careful spectral analysis, this physical interpretation lacks clarity mainly because the expected concentrations of perturbed molecules are too high compared to those we observe (Fig.~\ref{fig5}). For example, in the 1:1 mixture of light and heavy water, one expects that half of the bonds with neighboring molecules are formed by deuterons, while the residuals are too small to be explained in this way. We believe our approach to the interpretation of the IR spectra of D$_2$O/H$_2$O mixtures by introduction of the ionic species is more suitable.

\subsection*{Ion concentrations}
\begin{table}
\caption{\label{tab4}Parameters of the Lorentz fit according to Eqs.~\eqref{ds1} and \eqref{ds2} to the experimental spectra shown in Fig.~\ref{fig5} and the ion concentrations $n_i$ for different molar fractions $f$, as calculated using Eq.~\eqref{intgrl}: $I$ is the intensity, $w$ is the half-width and $\nu_2$ is the central frequency.}
\begin{tabular}{c|cccccc}
~ & $I$ & $w$ & $\nu_2$ & $n_i$ & $n_i$ & $n_i$\\
~ & - & (cm$^{-1}$) & (cm$^{-1}$) & $f=1/4$ & $f=1/2$ & $f=3/4$\\
\hline
H$_2$O & 15.0	& 50 & 1663 & 55 & 24 &6\\
~ & ~ & ~ & ~ & ~ & ~ &\\
D$_2$O & 13.5	& 27 & 1213 & 6 & 24 & 55\\
~ & ~ & ~ & ~ & ~ & ~ &\\
H$_3$O$^{+*}$ & 3.1 & 40 & 1695 & 0.53 & 0.16 & 0.02\\
~ & ~ & ~ & ~ & ~ & ~ &\\
DH$_2$O$^{+*}$ & 3.1 & 40 & 1637 & 0.53 & 0.47 & 0.18\\
~ & ~ & ~ & ~ & ~ & ~ &\\
HD$_2$O$^{+*}$ & 2.3 & 27 & 1222 & 0.18 & 0.47 & 0.53\\
~ & ~ & ~ & ~ & ~ & ~ &\\
D$_3$O$^{+*}$ & 2.3 & 27 & 1195 & 0.02 & 0.16 & 0.53\\
~ & ~ & ~ & ~ & ~ & ~ &\\
\end{tabular}
\end{table}

Having understood the origin of the discrepancy of the ``pure'' HDO spectra obtained at different molar fractions $f$, we can calculate the concentration of ionic species in water from their S-features. The spectral weight S of a Lorentz oscillator is related to the density of charges involved in the absorption process \cite{ref32,ref32a}:

\begin{equation}\label{intgrl}
S_i = \int_0^{\infty} \sigma(\omega'){\rm d}\omega' = \frac{\displaystyle \pi}{\displaystyle 2}\frac{\displaystyle n_iq_i^2}{\mu_i}
\end{equation}

\noindent where $\mu_i$, $q_i$ and $n_i$ are the effective mass, charge, and concentration of the conducting species of the $i$-th kind. Taking into account the effective masses and the parameters of Lorentzians obtained above, one can calculate the concentrations of the corresponding water species. The results are summarized in Table~\ref{tab4}. The concentrations obtained are in good agreement with the expected percentage for the different species in liquid water presented in  Table~\ref{tab2}. Thus, we have shown that the subtraction residuals shown in Figs.~\ref{fig1}B,C correspond to the unaccounted ionic species of H$_3$O$^{+*}$, DH$_2$O$^{+*}$, HD$_2$O$^{+*}$, and D$_3$O$^{+*}$.

\section*{Discussion and concluding remarks}
We have studied the infrared spectra of the H$_2$O, D$_2$O, and their mixtures in the direct transmission mode to uncover the hidden IR vibrations of water's ionic species. Our study reveals that the intramolecular vibrational dynamics observed in the IR spectrum of water is not reduced to the oscillatory motion of molecular species only. The presence of the residues on the spectrum of ``pure'' HDO, obtained from the mixtures with different molar fractions of light and heavy water, clearly shows that absorption of IR waves by water is partially defined by the dynamics of ionic species, whose concentration (at least at short times $<0.1$ ps – the low frequency limit of TIR) is about 2\% of the total number of molecules in liquid water. In order to bring this result in accord with pH concept, we suggest that fluctuation-driven short-living ions coexist with conventional long-living pH-active ions at short times of observation corresponding to the IR frequency range. The detected ionic species are important for water-related processes, e.g. solvation, dissolution, osmosis, cavitation, radiolysis, which take place on short length (nanometer) and time (sub-picosecond) scales \cite{ref33,ref34,ref34b}. We believe that an account for the ionic species on the ultrashort time scale can help to improve, and significantly simplify models of physical-chemical and electrochemical systems where liquid water plays a significant role \cite{ref35,ref36}. The space-time intervals discussed in our work are suitable for ab-initio approaches; and molecular dynamics simulations are highly welcome.

\section*{Methods}

Samples of H$_2$O and D$_2$O with purity of 99.9\% obtained from Sigma-Aldrich were used in this work. Probes with different molar fractions of deuterium, $f$, were prepared by mixing pure light and heavy water before measurements. The spectra were recorded in transmission mode \cite{ref24} using a Bruker Vertex 80v spectrometer in the frequency range 600-8000 cm$^{-1}$. For these measurements, a drop of water was sandwiched between two 3-mm-thick ZnSe windows. ZnSe is perfectly suited for such a type of measurements: it is hydrophobic and highly transparent with no absorption lines in the range of interest. There were no spacers between the windows. Instead, we controlled the stabilization of the water layer, measuring the IR spectra every 30 seconds until they were stabilized. The thickness for the light, heavy, and semi-heavy water samples was about 1 $\mu$m, slightly varying from sample to sample. The reference spectrum of the windows was recorded separately in advance. In order to eliminate the multiple reflections inside the water layer, the spectra were processed using a five-media model \cite{ref24} and an algorithm of least-squares minimization \cite{ref25}.

For our study, we intentionally chose the transmission mode (TIR) instead of the attenuated total reflection (ATR) technique \cite{ref26}. Although the ATR method provides the spectra without any fringes caused by multiple reflections within the water layer, the true absorption spectrum with the absolute values of transmitted intensity is only available in the TIR. As there is a discrepancy between the liquid water spectra obtained by these two methods (Fig.~6 of the Appendix), for the spectral weight analysis and the calculation of the ion concentrations a directly obtained TIR spectrum was used.

\section*{Acknowledgements}
We thank Prof. Dr. Joachim Maier for fruitful discussions. The work was partially supported by the Deutsche Forschungsgemeinschaft (DFG) via DR228/61-1. E.U. acknowledges the support by the European Social Fund and by the Ministry for Science, Research and the Arts of Baden-W\"urttemberg.

\newpage

~\\

\appendix
\section*{Appendix: Additional spectra and data}

\begin{figure}[h!]
\includegraphics[width=0.5\textwidth]{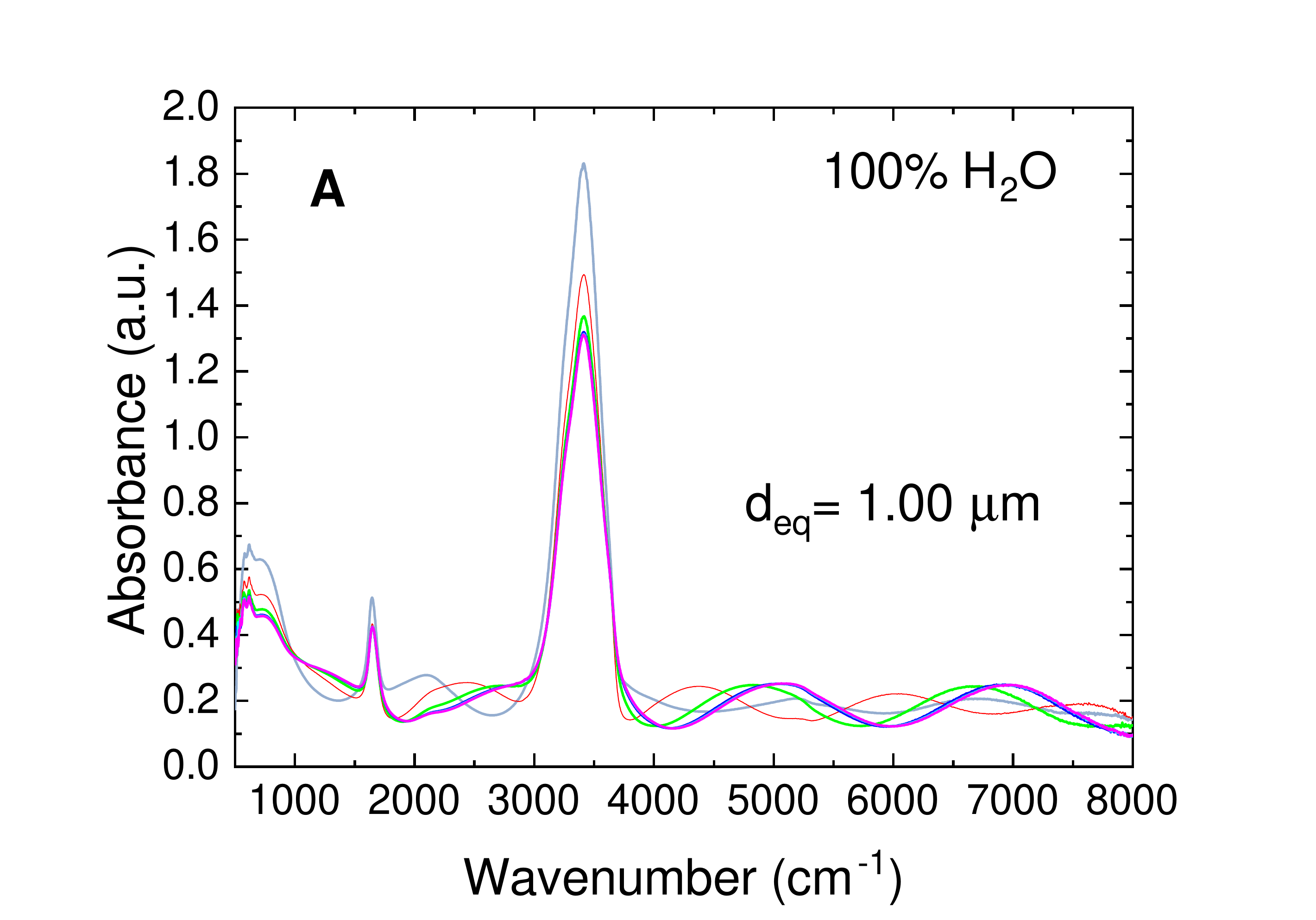}
\includegraphics[width=0.5\textwidth]{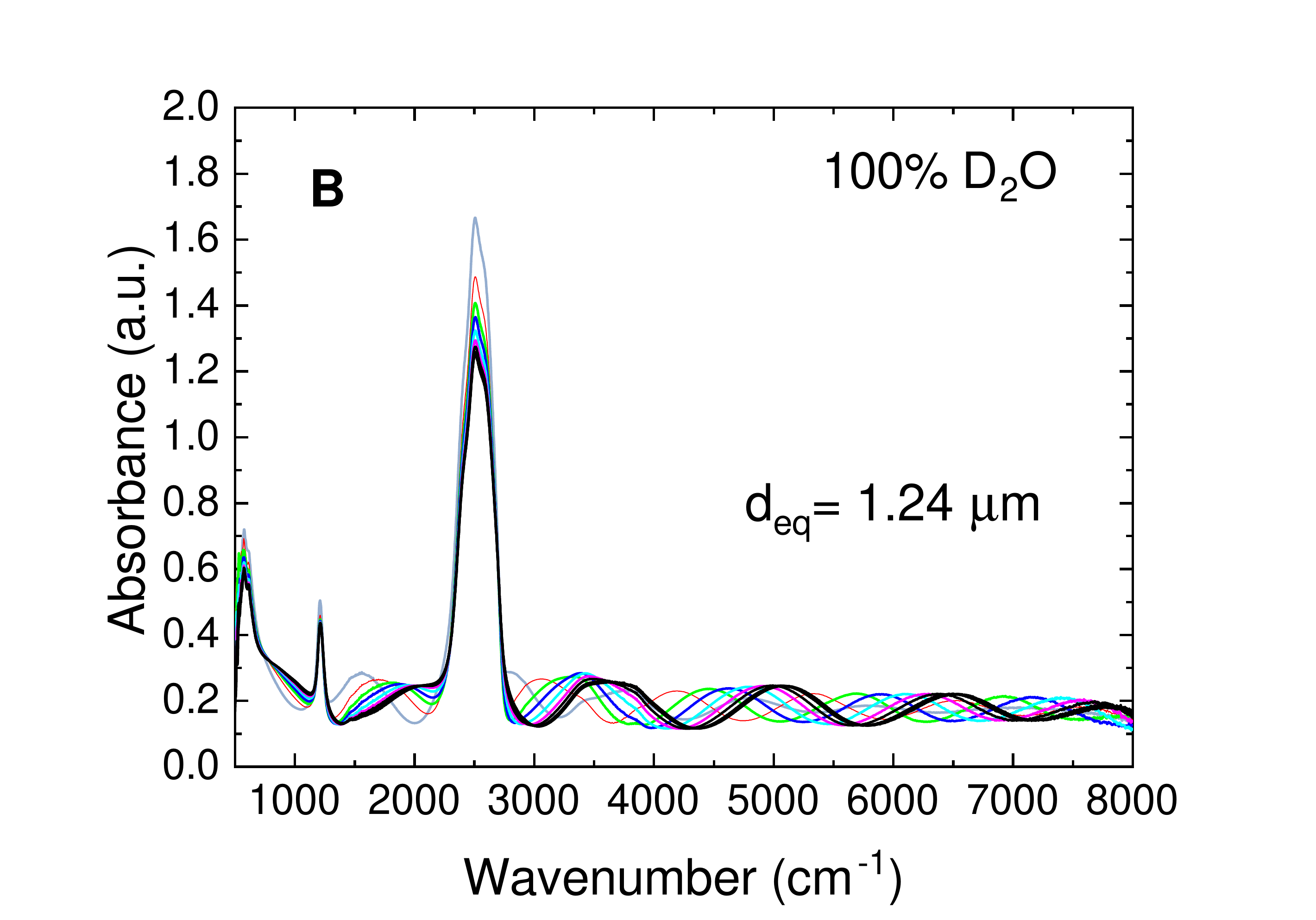}
\caption{Sets of absorption spectra of water squeezed between two ZnSe windows in infrared frequency region: (A) pure H$_2$O, (B) pure D$_2$O. The time interval between the measurements is 30 seconds. The parameter $d_{\rm eq}$ is the effective water layer thickness at equilibrium.}\label{figS1}
\end{figure}

\newpage 

\begin{figure}[h!]
\includegraphics[width=0.5\textwidth]{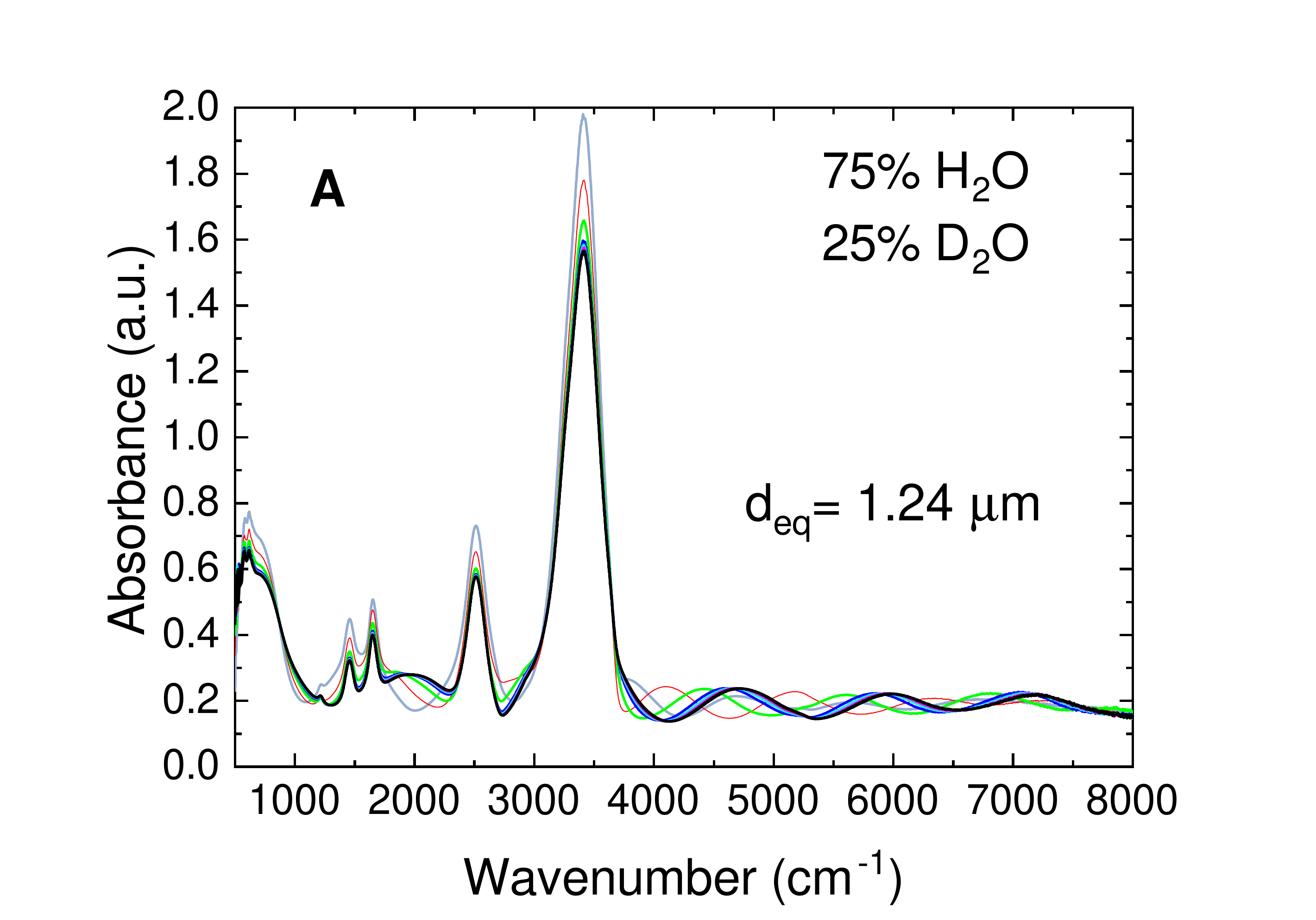}
\includegraphics[width=0.5\textwidth]{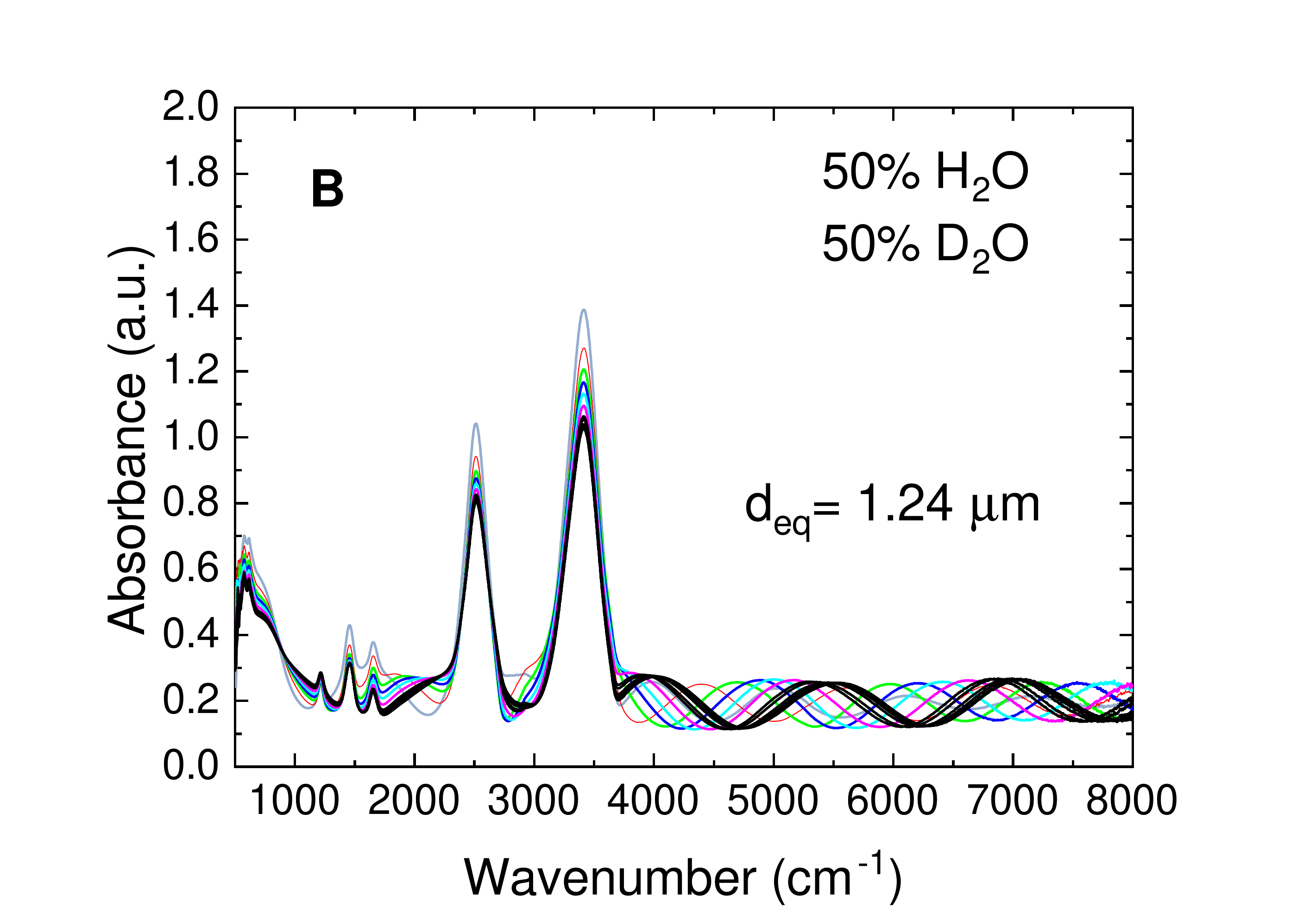}
\includegraphics[width=0.5\textwidth]{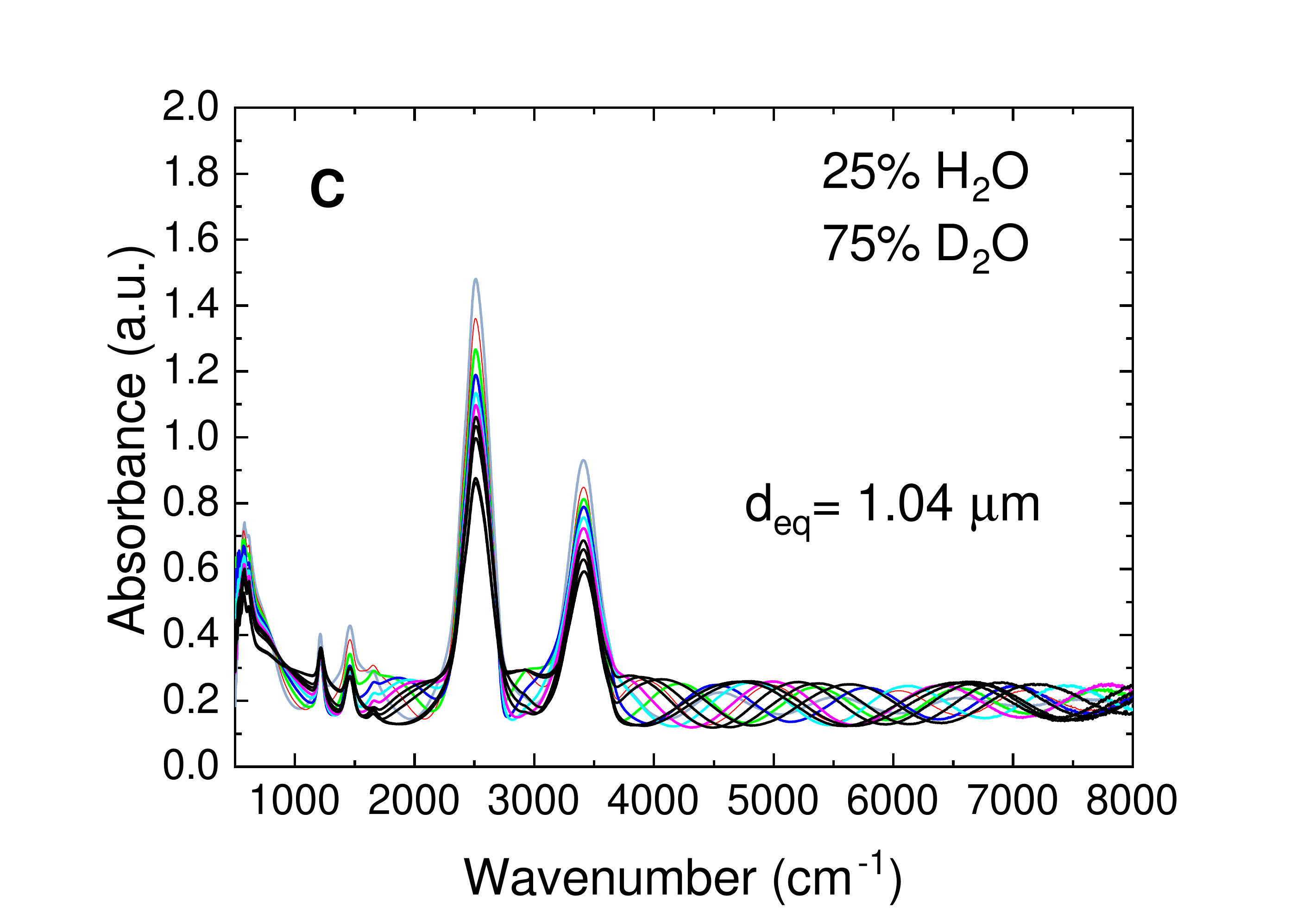}
\caption{Same as in Fig.~\ref{figS1} for light and heavy water mixtures: (A) 75\% H$_2$O and 25\% D$_2$O, (B) 50\% H$_2$O and 50\% D$_2$O, (e) 25\% D$_2$O and 75\% H$_2$O.}\label{figS2}
\end{figure}

\newpage 

\begin{figure}[h!]
\includegraphics[width=0.5\textwidth]{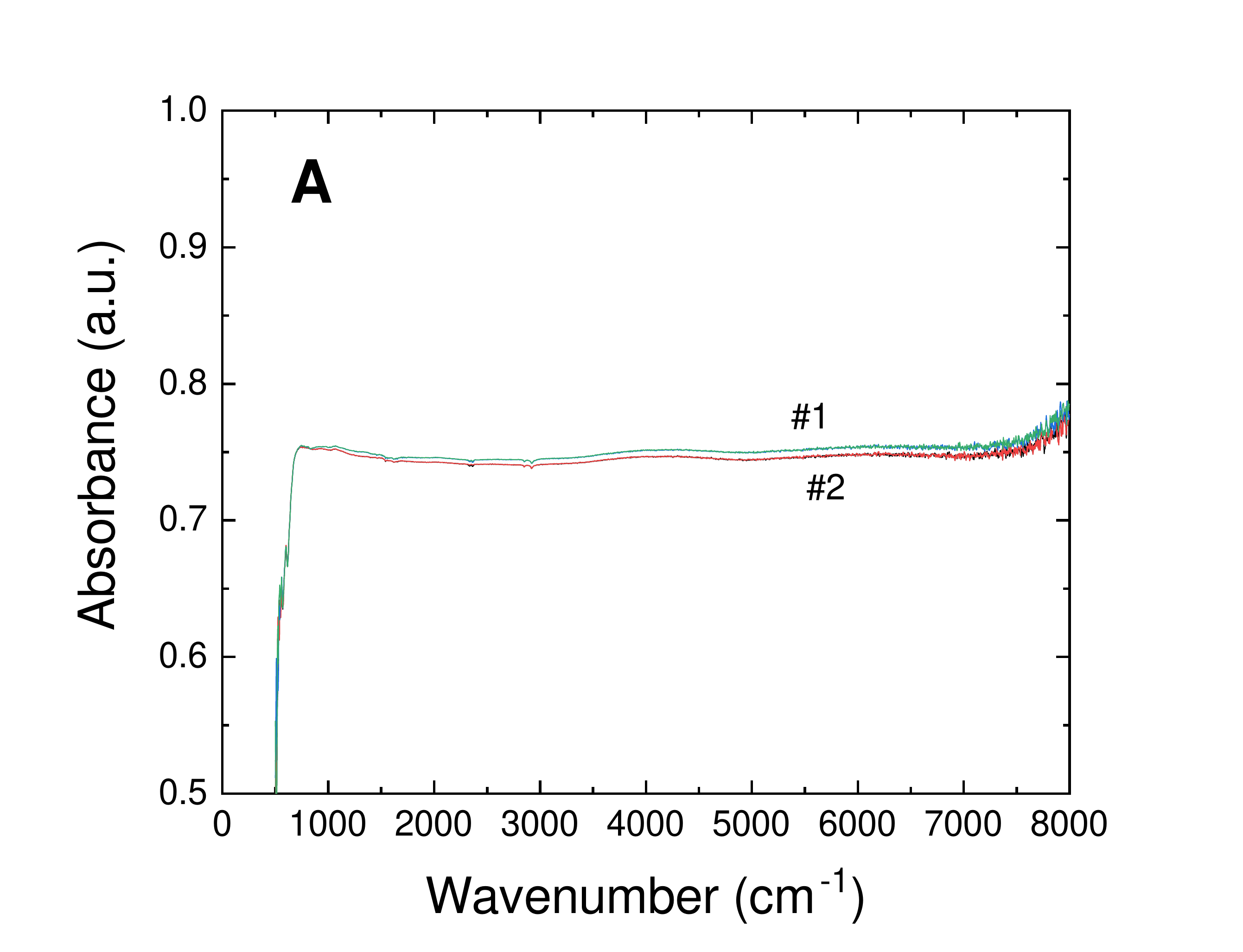}
\includegraphics[width=0.5\textwidth]{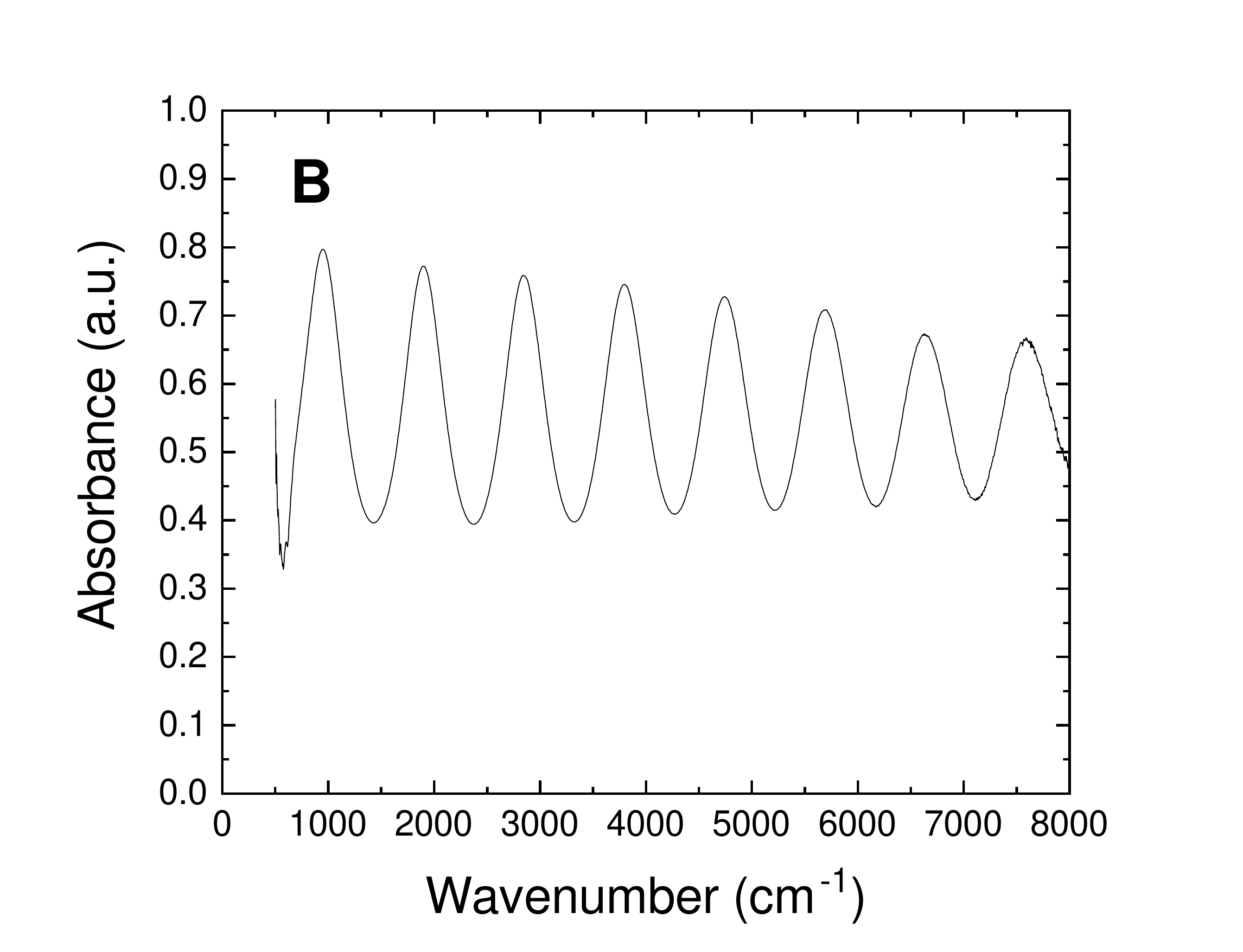}
\caption{Spectra of optically polished ZnSe windows used for the measurements: (A) separately; (B) two windows together. The thickness of each window is 3.0 $\pm$ 0.1 mm.}\label{figS3}
\end{figure}

\newpage 

\begin{figure}[h!]
\includegraphics[width=0.5\textwidth]{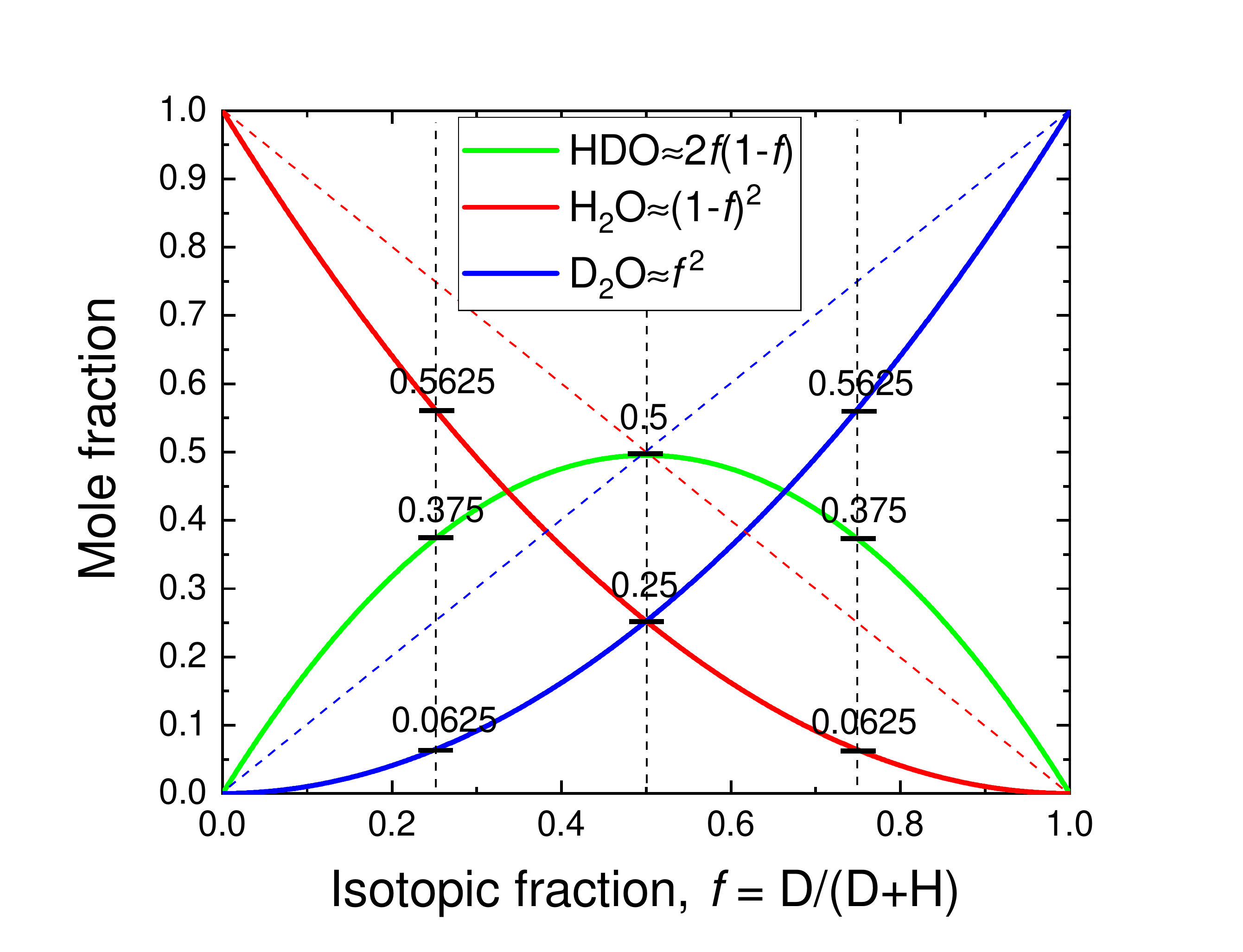}
\caption{Molar fractions of the components of light and heavy water mixtures as a function of molar isotopic fraction $f$. The curves are obtained with the equation $K=[\mbox{HDO}]^2/[\mbox{H}_2\mbox{O}][\mbox{D}_2\mbox{O}]\approx 3.85$ and the formulas given in the legend. Numbers near the curves are for $f = 1/4, 1/2, 3/4$ - the molar fractions of the mixtures, whose spectra are shown in Fig.~\ref{figS2}.}\label{figS4}
\end{figure}

\begin{figure}[h!]
\includegraphics[width=0.5\textwidth]{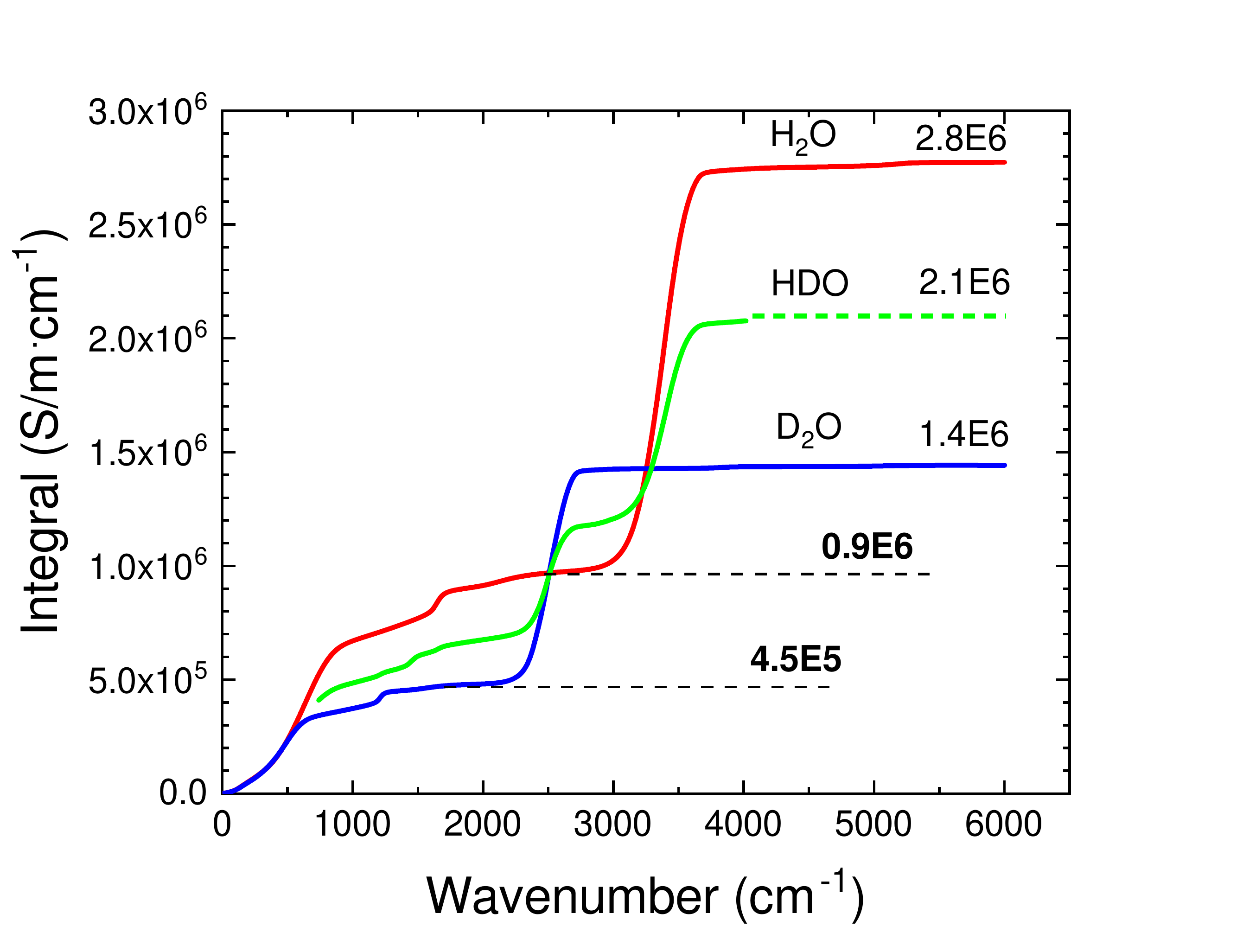}
\caption{Partial integrals of the spectra shown in Fig.~1, in the main text.}\label{figS5}
\end{figure}

\begin{figure}[h!]
\includegraphics[width=0.5\textwidth]{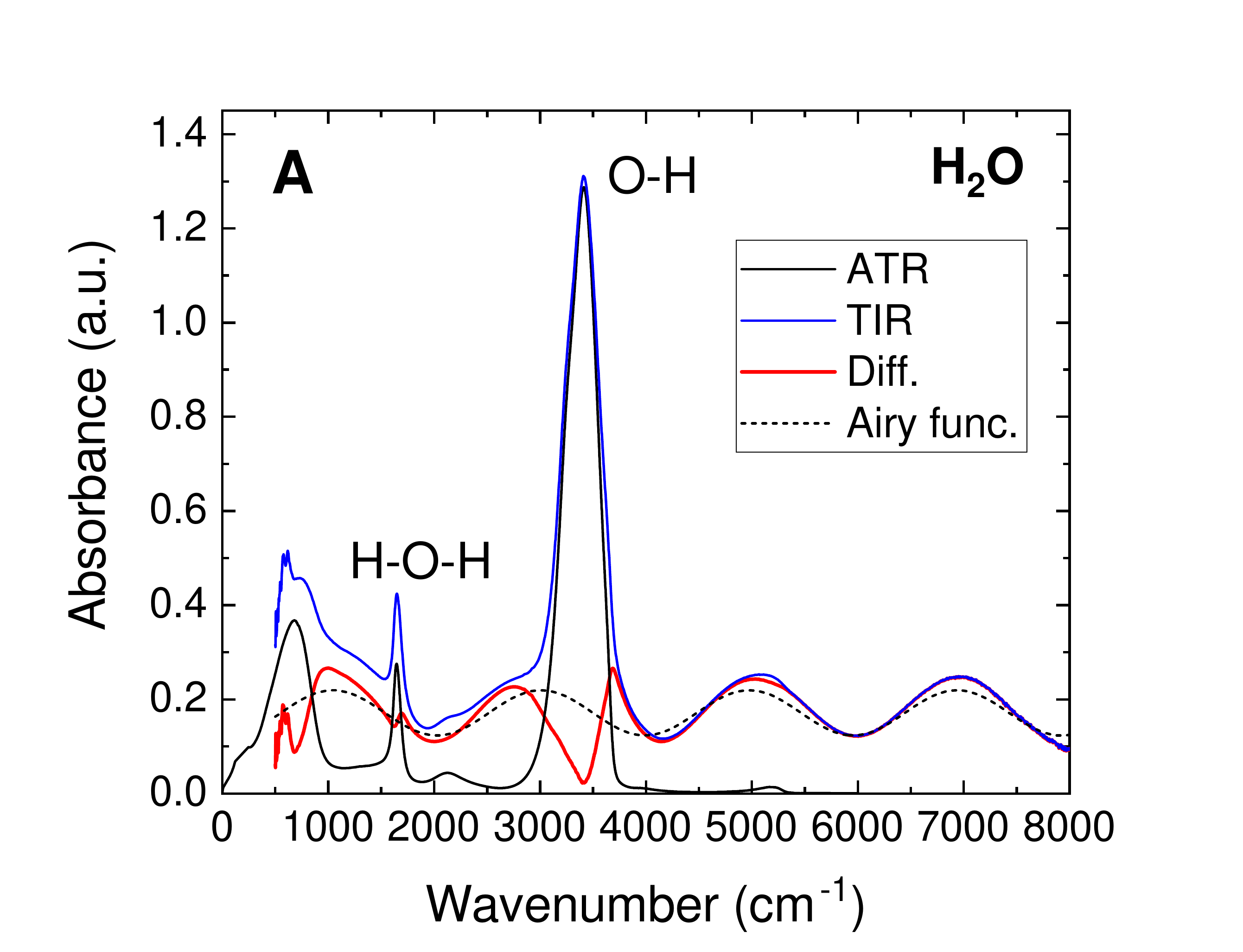}
\includegraphics[width=0.5\textwidth]{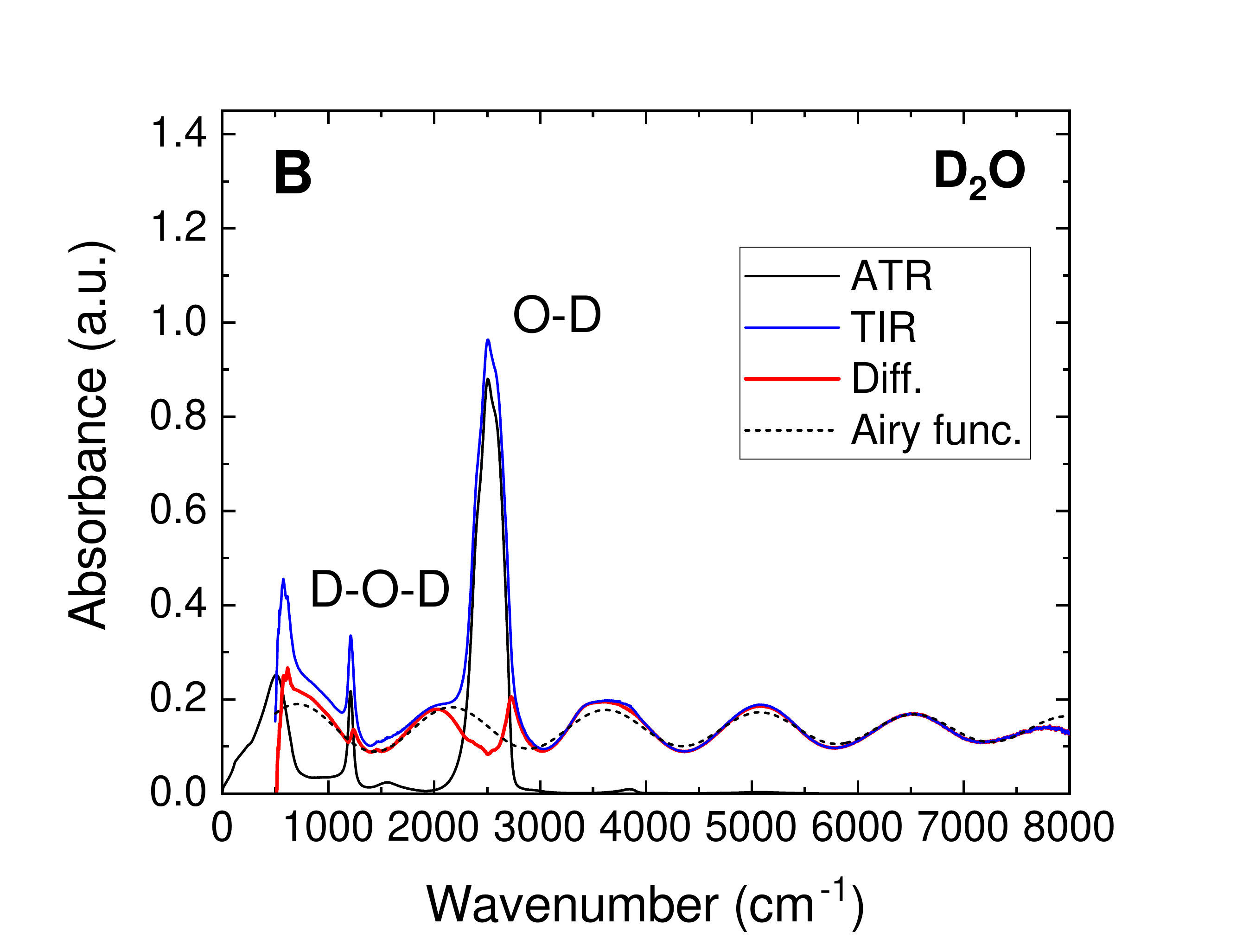}
\caption{Comparison of ATR (attenuated total reflection - black) and TIR (direct transmission - blue) spectra by subtraction (red). One expects that the difference should be an Airy function, but the characteristic deviations are observed near the stretching and bending modes. The ATR data are from Ref.~\cite{ref29}.}\label{fig6}
\end{figure}

\newpage 

\end{document}